# Towards Improving the NIST Fingerprint Image Quality (NFIQ) Algorithm

(Extended Version)

Johannes Merkle*, Michael Schwaiger*, Oliver Bausinger†, Marco Breitenstein*, Kristina Elwart†, Markus Nuppeney†

**Abstract:** The NIST Fingerprint Image Quality (NFIQ) algorithm has become a standard method to assess fingerprint image quality. However, in many applications a more accurate and reliable assessment is desirable. In this publication, we report on our efforts to optimize the NFIQ algorithm by a re-training of the underlying neural network based on a large fingerprint image database. Although we only achieved a marginal improvement, our work has revealed several areas for potential optimization.

This article represents the full version of [Me10], copyright of Gesellschaft für Informatik (GI).

## 1 Introduction

Assessing the quality of fingerprint images is crucial for many biometric applications. In particular, in the enrollment process it must be ensured that the provided fingerprint images are of sufficiently high quality. In the context of biometric applications, the quality is defined by the expected recognition accuracy (in terms of low error rates) in fingerprint verification with respect to this image. Consequently, the quality of a fingerprint image depends on the feature extraction and template comparison algorithms used. A lot of fingerprint verification software comes with a specific quality assessment algorithm for fingerprint images, which aims at predicting the recognition accuracy of the images when using the feature extractor and template comparison algorithm included in the software package. However, in applications where the fingerprint verification software of various manufacturers can be deployed, such a vendor-specific quality assessment is of limited use and a vendor-independent quality measure is needed.

In August 2004, NIST issued the NIST Fingerprint Image Quality (NFIQ) algorithm as part of the NIST Biometric Image Software (NBIS) [WGT+06]. The NFIQ algorithm is an open source tool for measuring the quality of fingerprint images independent of the fingerprint verification software used. The NFIQ measures quality by 5 classes, where class one refers to "excellent" and class five to "poor", and the "NFIQ value" output by NFIQ algorithm refers to the class number of the input fingerprint. As explained in [TWW04], the NFIQ algorithm is based on an artificial neural network that tries to predict the quality class from 11 features of the image. These features include the numbers of minutiae and image blocks with quality index exceeding several thresholds. The neural network has been trained with a large number of fingerprint images and the corresponding comparison statistics obtained with different fingerprint verification software.

\* secunet Security Networks AG, D-45128 Essen, Germany
† Bundesamt für Sicherheit in der Informationstechnik, D-53175 Bonn, Germany

While the present NFIQ algorithm is extremely useful, there still is demand for optimization. First, the division in five quality classes is quite rough: empirical tests reveal that for typical fingerprints more than 45% have NFIQ class 1. Certain applications, e.g. template protection techniques, may require particularly high quality images that can not be recognized by the NFIQ. Second, practical experience shows that the accuracy of the assessment is limited. On the other hand, an evaluation of the methodology followed for the development of the NFIQ algorithm [TWW04] indicates that there is still room for improvements: in particular, 40% of the fingerprint images used for the training of the neural network were not obtained from live scans but from inked impressions; clearly this type of fingerprint image is not relevant for authentication systems based on electronic capture devices. Furthermore, the training set consisted only of two imprints per finger, which rendered the estimation of the genuine matching performance difficult and inherently imprecise.

In a first attempt to improve the NFIQ algorithm, we conducted a complete re-training of the neural network based on a large database of live scan fingerprint images. Although the main focus was to improve the significance of the quality assessment for recognition performance by using a better data basis, we detected and implemented other optimizations of the NFIQ training. Unfortunately, evaluation of the resulting fingerprint quality assessment algorithm (subsequently denoted as NFIQ+) showed only marginal improvement with respect to error rates when compared to the original NFIQ algorithm. Nevertheless, our investigations conducted in the context of this re-training have revealed promising optimization potential that should be evaluated and exploited in future projects.

The present article is an extended version of the paper published at the BIOSIG conference [Me10].

This document is structured as follows. In Section 2 we outline the approach followed by NIST to implement the original NFIQ algorithm. In Section 3 we describe our efforts to re-train the neural network to obtain an improved NFIQ+ algorithm. In Section 4 we report the evaluation results of the NFIQ+ algorithm and its comparison to the original NFIQ. Eventually, in Section 5 we summarize the lessons learned and point out the potential for future optimizations.

## 2 Training of the Original NFIQ Algorithm

According to [TWW04], NIST used over 40.000 fingerprint images from 7 different databases for the development of the NFIQ and evaluated the similarity score statistics with 14 fingerprint software development kits (SDK). However, the neural network training was performed on a subset of 3900 fingerprints and using only 3 SDKs from Cogent, Sagem and NIST. Of these 3900 fingerprints in the training set, approximately 40% were scanned inked imprints, comprising both rolled and plain impressions. Furthermore, the training set contained two imprints per finger.

Subsequently, we use the following notations: we denote the fingerprints in the training set by $x_{f,i}$, where $f$ enumerates the fingers (we do not distinguish persons but only the fingers) and $i$ enumerates the imprints of this finger. For fixed SDK, we denote the similarity scores output by the comparison operation of $x_{f,i}$ and $x_{f',i'}$ by $S(x_{f,i}, x_{f',i'})$.

The approach of NIST was to determine a quality measure (the NFIQ) of the fingerprints by statistical evaluation of their similarity scores, and then to use these (statistically determined) NFIQ values for a training of a neural network that aims at predicting the NFIQ from certain image features representing relevant characteristics of the fingerprint. The overall process can be summarized as follows:

- For each pair of fingerprints in the training set, a comparison operation was conducted with each SDK; for each comparison operation the SDK outputs a similarity score indicating the level of similarity between the compared fingerprints. The matrix of the similarity scores resulting from all comparison operations with a specific SDK is called *similarity matrix*.

- For fixed SDK, the quality of each fingerprint was measured by (bins of) a function called *normalized match score* that depends on all its similarity scores. The normalized match score was defined as

$$o(x_{f,i}) := \frac{s_m(x_{f,i}) - \mu_n(x_{f,i})}{\sigma_n(x_{f,i})}, \tag{1}$$

where $s_m(x_{f,i}) = S(x_{f,i}, x_{f,i'})$ was the (only) genuine (match) score of $x_{f,i}$ in the test set, and $\mu_n(x_{f,i})$ and $\sigma_n(x_{f,i})$ denote the mean value and standard deviation of its impostor (non-match) scores, i.e. the $S(x_{f,i}, x_{f',i'})$ with $f \neq f'$. This definition is based on the assumption that a good fingerprint yields high genuine scores, low impostor scores, and a small deviation in impostor scores. The normalized match scores were computed for each SDK individually.

- The fingerprints were binned into five classes (the NFIQ) according to their normalized match scores, where class one covers the fingerprints with highest score and class five those with lowest score. The bins are defined by unevenly selected quantiles of the frequency distribution of the normalized match score. By selection, the training set contained only fingerprints where the class is unambiguous over all SDKs, i.e. fingerprints for which the class varied between the considered SDKs were excluded from the training set.

- For all fingerprints, 11 *features* were computed using the NBIS (NIST Biometric Image Software) package [WGT+06]. These features were supposed to represent the characteristics of the image relevant for its recognition performance, and were based on the quality values computed by the MINDTCT feature extraction algorithm (which is part of the NBIS package) for minutiae and image regions (blocks). In particular, the features contained

    ○ the number of pixels covered by the fingerprint in the image,

    ○ the total number of minutiae detected in the fingerprint,

    ○ for several thresholds, the number of minutiae with quality value exceeding this threshold, and

    ○ for several thresholds, the number of blocks with quality value exceeding this threshold.

    The quality values of minutiae and blocks are computed from three maps created during feature extraction marking areas with low contrast, without dominant ridge flow, or with high curvature of the ridge flow. These properties are assumed to reduce the reliability of detection of minutiae and other local fingerprint features (e.g. ridge flow orientation) utilized by comparison algorithms.

- A neural network was trained with the features and NFIQ classes of the fingerprints in the training set. In particular, a 3-layer feed forward perception network was used that takes as input the 11 features of the fingerprint and outputs an approximation of the NFIQ class. Using the real NFIQ (determined from the normalized match scores) the deviation from the correct answer was computed and the internal weights of the neural network were corrected accordingly.[1] A second set of fingerprints along with their (statistically determined) NFIQ and image features was used to continuously check the prediction accuracy and to stop training as soon as the accuracy was settled; well-timed ceasing of the training is crucial to prevent over-specialization of the neural network with respect to the training set.

Particular attention was given to the treatment of the genuine scores, as they may not equally depend on the quality of both fingerprints. In particular, in [TWW04] the plausible hypothesis was established that the genuine score depends on the fingerprint having lower quality; for instance, the comparison of a fingerprint of high quality and a poor quality imprint of the same finger will result in a low score. As a consequence, a genuine score is hardly significant for assessing the quality of the better quality fingerprint, at least, if the quality of both fingerprints differ considerably. For this reason, NIST used the trained neural network to assess for each finger which of the two fingerprints has higher quality and to perform a second training of the neural network, where to each fingerprint a *pattern weight* was assigned, which determined its influence on the network training:

- for fingerprints that were assessed to be the higher quality imprint of this finger a pattern weight of zero was assigned, implying that it had no influence to the training;

- for fingerprints that were assessed to be the lower quality imprint of this finger a pattern weight of one was assigned, implying that it had full influence to the training;

- if both imprints of a finger were assessed to be of equal quality, both were assigned with pattern weight 0.5.

---

1  The neural network and the training algorithm deployed were originally developed and deployed for fingerprint pattern classification by the PCASYS algorithm in the NBIS package [WGT+06].

Many more details about this process can be found in [TWW04]; the most important ones for optimization will be discussed in Section 3.

## 3 Re-Training the Algorithm (NFIQ+)

### 3.1 Fingerprint Images and SDKs Used

We used a database consisting of 9 imprints of 8784 fingers from 1098 individuals. The fingerprints were taken with 3 different optical sensors, each of which used to capture 3 images per finger. Since the sensor significantly influences the image quality, using a sufficient variety of sensors is crucial to avoid overspecialization of the NFIQ+. We used the imprints of 4392 fingers for training and the other 4392 fingers for testing.

We used 5 SDKs from Dermalog, L1, Neurotechnologija, NIST, and NEC (in alphabetical order). For the sake of anonymity, the algorithms will subsequently be referred to as SDK A-E with random order. For all SDKs, the same fingerprint comparison operations were computed resulting in according sets of similarity scores.

Since the computation of a complete similarity matrix of the 39528 fingerprints would take too much time, we computed (for each SDK) only 450 similarity scores for each fingerprint, 442 of which were impostor scores with imprints of 442 different fingers and 8 were the genuine scores with the imprints of the same finger. We computed impostor scores only between fingers of different individuals, because comparisons between fingers of the same person may yield different score distributions (see Section 8.2.4.5 in [ISO1]).

### 3.2 Treatment of Genuine Scores

The computation of 8 genuine scores per fingerprint (evaluation of the scores was conducted for each SDK separately) allowed an accurate consideration of the genuine scores as compared to the development of the original NFIQ, where only a single genuine score was available. .

Obviously, multiple samples allow a more accurate estimation of the expectation value. However, following the hypothesis of [TWW04] a genuine score is mainly significant for the lower quality fingerprint. With several genuine scores per fingerprint available, we were able to á priori (i.e. prior to the neural network training) rank the imprints of a finger according to their quality. We could then use this (preliminary) ranking to identify which genuine scores are significant for which fingerprints and finally could compute the normalized match score considering only the relevant (i.e. significant for the fingerprint) genuine scores. In particular, we implemented the following process:

1. For each fingerprint we computed the average over all its genuine scores, and used this value to compute a preliminary normalized match score and its corresponding preliminary rank among all fingerprints (over all fingers).

2. We used this ranking to identify for each fingerprint the set of *significant* genuine scores, which were all genuine scores with imprints of this finger having at least approximately the same quality. Precisely, we regarded only the scores with those fingerprints, whose preliminary rank was at most 1/6 below the rank of the fingerprint in question.

3. We re-computed the normalized match scores, this time (for each fingerprint) considering only the significant genuine scores.

In the third step, the number of significant genuine scores could be quite small, in extreme cases even zero. This happened in particular for fingers where the quality of the 9 fingerprints varied significantly. In such a case, special treatment of this fingerprint was necessary, as described in Section 3.3.

As a second optimization potential, we were able to measure the dispersion of the genuine scores. It is clearly desirable that a fingerprint does not only produce high average genuine scores but also yields these high scores consistently over all imprints of comparable or better quality. Therefore, we modified the definition of the normalized match score as described in Section 3.3.

### 3.3 Defining the Normalized Match Score

The definition of NIST for the normalized match score (1) was based on the limitation that only one genuine score is available per fingerprint. Therefore, we had to generalize it to the case of several genuine scores by replacing the (only) genuine score $s_m(x_{f,i})$ in (1) by the mean of the genuine scores that are significant for this fingerprint (see previous section for a discussion). As a result, we obtained the following function

$$o_1(x_{fi}) := \frac{\tilde{\mu}_m(x_{f,i}) - \mu_n(x_{f,i})}{\sigma_n(x_{f,i})} \qquad (2)$$

where $\tilde{\mu}_m(x_{f,i})$ denotes the mean of the genuine scores that are significant for fingerprint $x_{f,i}$.[2]

On the other hand, as argued in Section 3.2, it is reasonable to regard the dispersion of the genuine scores as well. An obvious approach would be to use the standard deviation of the significant genuine scores of a finger in the modified function $o_2(x_{fi})$ for the normalized match score, e.g, by replacing the denominator in (2) with $\sigma_n(x_{f,i}) + \tilde{\sigma}_m(x_{f,i})$ (as suggested in the annex of [ISO2]) or $\sqrt{\sigma_n(x_{f,i})^2 + \tilde{\sigma}_m(x_{f,i})^2}$, where $\tilde{\sigma}_m(x_{f,i})$ denotes the standard deviation of the significant genuine scores of fingerprint $x_{f,i}$. However, although there are robust estimators for the standard deviation from very small samples [RV02], we have to expect considerable inaccuracy in an estimation from at most 8 values. As the genuine scores are generally much greater than the impostor scores, this will result in significant noise in the denominator. Furthermore, the deviation of the genuine scores would predominate the deviation of the impostor scores in the denominator.

Therefore, we decided to choose a definition of the normalized match score, where the dispersion of the significant genuine scores (in particular, the deviation below the mean value) is represented in the nominator. We found such a definition, by replacing the mean of the significant genuine scores $\tilde{\mu}_m(x_{f,i})$ in (2) with an appropriate $\alpha$-quantile of the fingerprint's genuine score distribution with $\alpha < 0.5$. The selection of the quantile size $\alpha$ is a compromise between accuracy and statistical significance: A large quantile, e.g. $\alpha = 0.3$, will not sufficiently represent the dispersion, while a small quantile, say $\alpha = 0.05$, can hardly be estimated from at most 8 samples. As a (heuristic) compromise we set $\alpha = 0.15$ and arrived at the definition

$$o_3(x_{fi}) := \frac{Q_{0.15}(s_m(x_{f,i})) - \mu_n(x_{f,i})}{\sigma_n(x_{f,i})}, \qquad (3)$$

where $Q_{0.15}(s_m(x_{f,i}))$ denotes the threshold that is (expected to be) exceeded by 85% of all significant genuine scores of fingerprint $x_{f,i}$. The estimation of this quantile from at most 8 samples is still error-prone, but its value is usually relatively close to $\tilde{\mu}_m(x_{f,i})$. Hence, if the number of samples is not too small, say 4, the impact of the noise from the estimation is relatively mild.

For all fingerprints, where at least 4 significant genuine scores were available, we used its rank with respect to the normalized match score according to definition (3) as a basis for the NFIQ+ classes. In all other cases, i.e. in those cases, where the number of imprints of the same finger with at least comparable quality was less than 4, we used its rank according to definition (2) as a fall-back solution. The rank-based fusion accommodates the fact that the two functions $o_1(x_{fi})$ and $o_3(x_{fi})$ produce incomparable values.

---
2   We use the tilde on top of the symbol $\mu$ to distinguish from the mean value over all genuine scores.

## 3.4  Robust estimation of statistical measures

The estimation of the mean and standard deviation of a fingerprint's impostor score distribution (over random comparisons within the world's population) from 442 samples is no problem because this sample size is sufficiently large to ensure a quite accurate estimation by means of the arithmetic mean and the sample standard deviation.

For a fingerprint's genuine scores, the situation is different, as we have at most 8 samples and in many cases, in particular for high quality fingerprints, the number of significant genuine scores is much lower. In order to estimate the mean and 15%-quantile of the significant genuine scores of a fingerprint from such small samples, we need robust estimators for these statistical measures. As shown in [WW05], the distribution of genuine and impostor scores is not Gaussian and even varies from vendor to vendor, and thus, we cannot assume any specific distribution of the genuine scores. We decided to use the median as the mean value, and to estimate the median (which is the 50% quantile) and the 15% quantile of the significant genuine scores by the quantile estimator of Harrell and Davis [HD82], which has been shown to be particularly robust for small samples (see [DLP94]). The estimator computes an approximation of a quantile of an arbitrary distribution by a weighted sum $Q_\alpha = \sum_{i=1}^{n} w_{n,i} x_i$ of the samples, where the samples $x_1 \leq \ldots \leq x_n$ are ordered by size and the constants $w_{n,i}$ are given by

$$w_{n,i} = \frac{B(i/n, (n+1)\alpha, (n+1)(1-\alpha)) - B((i-1)/n, (n+1)\alpha, (n+1)(1-\alpha))}{B(1, (n+1)\alpha, (n+1)(1-\alpha))},$$

with

$$B(z, a, b) = \int_0^z y^{a-1}(1-y)^{b-1} dy$$

being the incomplete beta function.

## 3.5  Multi-Algorithmic Fusion and Definition of NFIQ+ Classes

For the training of the original NFIQ algorithm, NIST restricted the training set to those fingerprints that had the same NFIQ class for all 3 considered SDKs. However, the neglected fingerprints might have specific properties, which are the reason for their algorithmic-dependent performance and, hence, are not equally present in those fingerprints that exhibit similar performance for all SDKs. Consequently, the NIST approach for fusion may have resulted in an NFIQ that shows poorer performance on such fingerprints. Therefore, we decided to follow a different approach and to implement a fusion method for the quality measures obtained for the individual SDKs.

Since the similarity scores from different vendors have different ranges and distributions, the resulting values of the normalizes match score are incomparable. On the other hand, fusion on the basis of vendor-specific NFIQ+ classes seemed to rough. Therefore, we deployed a rank-based fusion: for each fingerprint, we computed the arithmetic mean of its ranks obtained with the individual SDKs, and then computed a final rank of all fingerprints over these mean values.

In order to evaluate the impact of the number and distribution of the NFIQ+ classes to the reliability of the prediction by the neural network, we used 3 different classifications:

- 10 classes over which the fingerprints are uniformly distributed
- 5 classes over which the fingerprints are uniformly distributed
- 5 classes for which the distribution of the fingerprints resembles that induced by the NFIQ algorithm.

For each of these class definitions, we trained the neural network as described in Section 3.6 and evaluated the results.

## 3.6 Image Feature Selection and Neural Network Training

Although we feel that the definition of the feature vectors bears great optimization potential, we decided, for the time being, to keep the feature vectors as they are. Selection of appropriate feature vectors is quite complex and should be conducted independently of other optimizations to allow a step-by-step evaluation of the impact of individual improvements. The feature vectors were computed using a dedicated tool that is included in the NBIS package (see [WGT+06]).(Of course, basing the feature extraction on the minutiae extractor of NIST may render the features more significant for the NIST matching algorithm than for the other SDKs.)

The neural network training was conducted using the training software in the NBIS package. This software executes the neural network on a set of feature vectors of fingerprints from the training set, and uses its outputs and the true class numbers to determine the error according to a configured error function. It then uses an optimization method to determine the changes to the internal network weights that would best reduce the error for the processed fingerprints. After each training run, an equally large number of data of fingerprints of an independent test set is processed to measure the prediction accuracy. The number of training runs can be configured, but the training can also stop as soon as the rate, at which the classes of fingerprints from the test set are correctly predicted, does not improve significantly.

The neural network and the training software have many parameters and alternative options, some of which are discussed below.

- **Number of nodes.** While the number of input nodes and output nodes are predetermined by the number of feature vector components and the number of NFIQ+ classes, respectively, the number of nodes in the hidden (i.e. middle) layer can be freely configured. In the original NFIQ training, NIST used 22 nodes in this layer.

- **Optimization method.** For optimization of the network weights with respect to the error, the NBIS training software supports the Scaled Conjugate Gradient (SCG, see [M93]), as well as the Broyden Fletcher Goldfarb Shanno (BFGS, see [NW06]) method.

- **Error function.** The error function measures the deviation of the output of the neural network, which is given be the activations[3] of the output nodes, from the perfect output, which would be an activation of one at the correct node (corresponding to the true NFIQ+ class of the input) and zero activation at all other output nodes. The NBIS package implements three different error functions, of which the Mean Squared Error (MSE) function is recommended by [WGT+06]. It measures the error of the activations $(x_1, \ldots, x_5)$ by

$$(1-x_i)^2 + \sum_{j \neq i} x_j^2,$$

where $i$ is the true NFIQ+ class. Since the NBIS neural network software was developed for classification problems and not for approximation of a function, the error functions neglect the distance of the individual output nodes from the correct node. However, the NFIQ+ is a gradual measure, and for a true NFIQ+ class 1 the activation of output node 2 is "less wrong" than the activation of node 5. Therefore, we modified the MSE error function so that it takes into account the distance between the individual output node and the correct node. In particular, we defined a error function MSE that computes the error as

$$\left((1-x_i)^2 + \sum_{j \neq i} |i-j| x_j^2 \right)/C. \qquad (4)$$

where $C$ is a constant that keeps the average error at the same value as with the original MSE.[4]

- **Regularization factor.** The regularization is used to limit the size of the network weights and to avoid over-fitting of the network to the training set.

---

3  As a result of the processing by the neural network each output node has an activation value which is normalized to range [0,1].

4  Larger average error values can result in global decrease of the network weights which can affect the prediction performance, in particular in the context with Boltzmann pruning.

- **Boltzmann pruning.** In order to improve the computational performance of the neural network, edges with very low weights can be removed. This pruning is performed in a random pruning process resembling thermodynamic processes using the *Boltzmann temperature* as a parameter.

# 4 Evaluation of Results

## 4.1 Inter-algorithmic Consistency of Score Statistics

In order to analyze the consistency of the statistical measures used for quality assessment between the deployed SDKs, we evaluated correlations. As shown in Table 1, the rank-based correlation of normalized match scores between all used comparison algorithms is predominantly consistent. The ranks calculated with the individual SDKs exhibit high correlation.

When using the definition (2) of the normalized match score, which is a direct generalization of the definition used by NIST, the results are equally consistent among the algorithms, i.e. the rank-based correlations of the normalized match scores among the SDKs were almost identical to the case of using definition (3).

The 15% quantile of the genuine scores are also very consistent between the different SDKs with correlation coefficients between 0.82 and 0.9. This high correlation is presumably the main reason for the consistency of the normalized match score. In contrast, the average impostor scores for the individual software packages are much less consistent, with correlations being predominately in the range between 0 and 0.36. However, as the impostor scores are typically much less than those of the genuine scores, in (3) the nominator is dominated by the 15% quantile of the genuine scores, and consequently the inconsistency of the average impostor scores should only have very limited impact to the normalized match scores.

|       | SDK A | SDK B | SDK C | SDK D | SDK E |
|-------|-------|-------|-------|-------|-------|
| SDK A | 1     |       |       |       |       |
| SDK B | 0.86  | 1     |       |       |       |
| SDK C | 0.75  | 0.75  | 1     |       |       |
| SDK D | 0.75  | 0.77  | 0.62  | 1     |       |
| SDK E | 0.61  | 0.59  | 0.70  | 0.48  | 1     |

*Table 1: Rank-based correlation of normalized match scores.*

Much worse for the inter-algorithmic consistency of the quality assessment is the standard deviation of impostor scores, which is the only term in the denominator of (3) and turned out to be quite inconsistent among the SDKs. As the values from Table 2 demonstrate, the rank-based correlation of impostor scores is very low in most cases. In the case of SDK D, the inconsistency is presumably due to fact that it internally applies a threshold and very frequently outputs the same minimal impostor score 0, which results in an unnatural low standard deviation of impostor scores. (Most matching algorithms perform impostor normalization internally but this process annihilates information needed for the training.)

|       | SDK A | SDK B | SDK C | SDK D | SDK E |
|-------|-------|-------|-------|-------|-------|
| SDK A | 1     |       |       |       |       |
| SDK B | 0.07  | 1     |       |       |       |
| SDK C | 0.43  | -0.10 | 1     |       |       |
| SDK D | -0.03 | -0.01 | -0.22 | 1     |       |
| SDK E | 0.38  | 0.14  | 0.69  | -0.19 | 1     |

*Table 2: Rank-based correlation of the standard deviation of impostor scores.*

## 4.2 Results of Re-Training

After training the neural network on 10 uniformly distributed classes, it predicted the NFIQ+ class correctly for 18.3 % of all fingerprints in our test set. This prediction accuracy was reached and remained stable already after just a few training runs.

Subsequently, we tested several optimizations of the neural network training to achieve better results in terms of predicting the correct class of the fingerprint imprint. Following list summarizes the outcomes of these attempts:

- *Variation of the regularization factor:* Mainly performance increases in re-training time were noticed by changing the default regularization factor of $10^{-4}$. The recognition performance did not change significantly.

- *Changing numbers of nodes in the hidden layer:* No increases of recognition performance was noticed by increasing and decreasing the default number of hidden layers (22).

- *Usage of BFGS optimization:* Using the BFGS method for optimization instead of the SCG method (see Section 7) resulted in a significant worse recognition performance. We also tried combining both methods (in a succession), but this approach did also not yield any improvement over using SCG only.

- *Adaption of class distribution:* Changing the class distribution to the distribution of the original NFIQ algorithm (non-uniformly distributed) leads to a recognition rate of about 48 % of the assigned NFIQ+ classification which is slightly higher as compared to the original NFIQ algorithm (46 % with regard to NFIQ+). Note that the increased prediction rate is not necessarily an improvement as 5 non-uniform classes are much easier to predict than 10 uniform classes. However, the resembling of the original NFIQ class distribution allowed a direct comparison by prediction rate with the original NFIQ algorithm, which was not possible with 10 classes.

- *Modification of the error function:* Out of three available error functions for training the neural network Mean Squared Error (MSE) returns the best results in terms of recognition rate. Our optimization of the error function according to (4) results in a marginal reduction of the rate at which the class is correctly predicted but also significantly reduces the frequency of larger deviations from the correct output. Table 3 summarizes the statistic of deviation of the predicted class from the correct value for the original MSE function and our optimized error function[5].

| Deviation of predicted classes from correct value | Frequency *MSE* | Frequency *Optimized MSE* |
|---|---|---|
| 0 (correctly predicted class) | 35.1 % | 34.7 % |
| 1 | 36.3 % | 41.1 % |
| 2 | 21.0 % | 19.9 % |
| 3-4 | 7.6 % | 4.3 % |

*Table 3: Deviation of predicted classes for original and optimized error function MSE.*

## 4.3 Biometric Performance Results

In order to determine the biometric performance of our re-trained NFIQ+ algorithm and to quantify the improvements compared to the original NFIQ algorithm, the following simulation was conducted using the test set of the available data basis (which is disjoint from the set used for training). For each finger, the "best" imprint in terms of fingerprint quality was selected using the quality assessment methods (e.g. the NFIQ and NFIQ+ algorithms) to be compared. (If quality assessment resulted in multiple imprints having the same quality, of these the imprint with lowest index in the database was selected.) The 450 similarity scores (442 impostor and 8 genuine scores) calculated for the statistical determination of the NFIQ+ classes (see Section 3.1) were then used to determine the false match rate (FMR) and false non-match rates (FNMR) of this imprint[6]. Results of all calculated error rates where then used to show the biometric performance with the help of a Detection Error Trade-Off (DET) curve. The same procedure was also applied to the original NFIQ algorithm.

---

5  5 uniformly distributed classes were used to obtain these values.

In order to evaluate how good the neural network predicts the NFIQ+ classes, we also calculated the error rates for the case that the best imprint is chosen from each triple according to the statistically determined classes. The DET curve of this imaginary *perfect class predictor* shows the optimal achievable error rates for the given statistical NFIQ+ definition. As a result, by comparing the DET curve of the NFIQ+ algorithm and the curve of this perfect class predictor, the prediction accuracy of the neural network can be assessed according to a very application-relevant metric.

The following figure shows DET curves for the re-trained NFIQ+ algorithm using 5 non-uniformly distributed classes and for the original NFIQ algorithm, as well as for the perfect class predictor (labeled "statistical classification") for 5 non-uniform classes. Unfortunately, the improvement by the re-training is quite small: when using the re-trained NFIQ+ algorithms, the equal error rate (EER) decreased to 6.38% from 6.50% for the NIST NFIQ algorithm. However, further improvement can be expected by using the modified error function (see Sections 3.6 and 4.2) which was not evaluated due to time constraints.

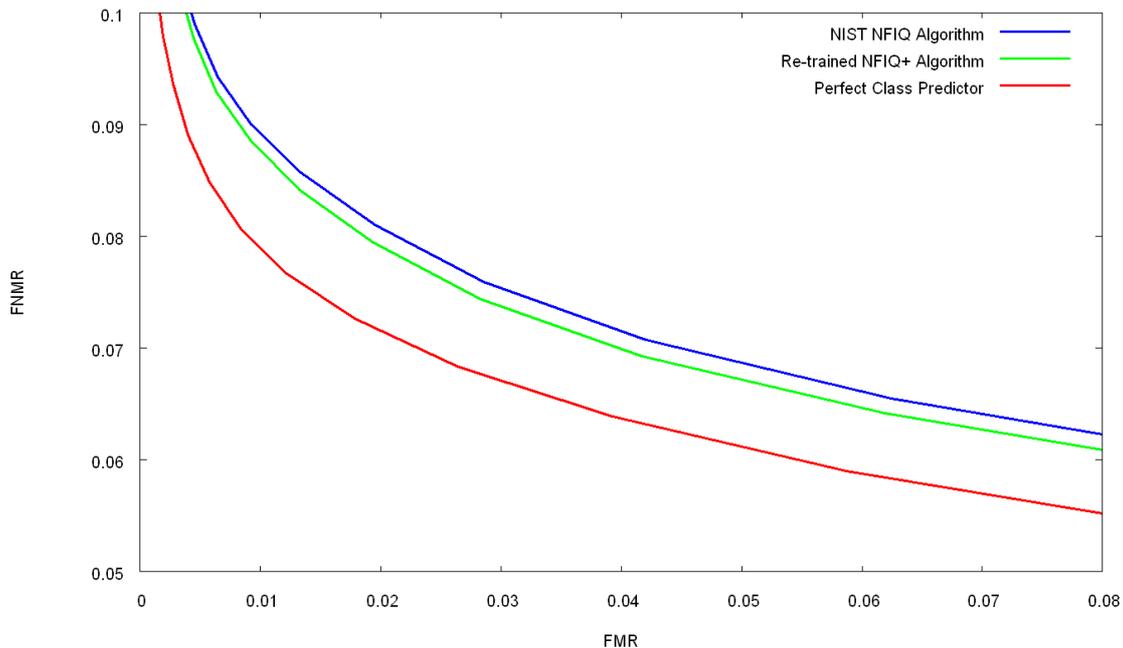

*Figure 1: DET curves of original and re-trained NFIQ algorithms for 5 non-uniform classes.*

By comparing the DET curves of the perfect class predictor for different class definitions (varying number and distribution of classes), we can determine which class determination is most promising with respect to biometric performance, i.e. bears the most potential with respect to error rates assumed that the class can be predicted with comparable accuracy by a trained neural network. We evaluated the DET curves for following three statistical classifications:

- 10 classes using uniform distribution,
- 5 classes using uniform distribution,
- 5 classes resembling the (non-uniform) distribution of the original NFIQ.

The result depicted in Figure 2 shows that the biometric performance is better for 10 classes than for 5 classes. On the contrary, the distribution of the classes seems to have relatively little impact on the biometric performance.

---

6  Bozorth3 from the NBIS package [WGT+06] was used as comparison algorithm.

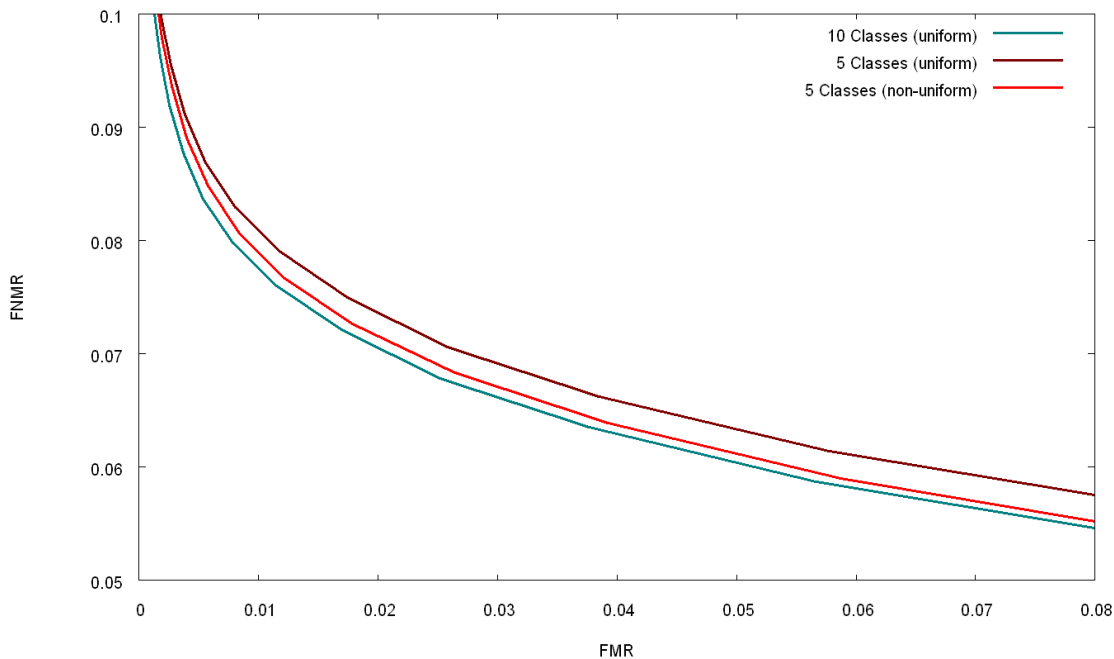

*Figure 2: DET curve for perfect class predictors of 3 different class definitions.*

The approach of evaluating the DET curves of the imaginary perfect class predictor for varying class definitions allows also assessment of other optimization potentials that influence the class definitions, e.g. the training set used, the statistical functions used, the fusion method, etc. We will discuss this approach in more detail in Section 5.

## 5  Conclusions and Future Work

In our re-training, the main focus was to investigate the optimization potential by a better data basis. Rolled and inked impressions, which were (among others) used in the original training of the NFIQ algorithm, are not relevant for authentication systems based on electronic capture devices. By using only plain live scans captured by optical fingerprint sensors the data basis was adopted for this purpose. Further optimization potentials arose from this new data basis. As more than 2 imprints per finger were available a more accurate and comprehensive treatment of genuine match scores was possible, which required a modification of the statistical functions used for quality assessment.

Furthermore, since 2004, new fingerprint software has been developed and are deployed. Thus, using up-to-date feature extractors and comparison algorithms makes the re-trained NFIQ algorithm fit for current applications.

Finally, we increased the number of classes as it seemed that the current categorization by the NFIQ is not exact enough and does not sufficiently differentiate good fingerprints. Conducting a rank-based fusion of all comparison algorithms was also a new approach considered within this re-training process.

Due to limitations in time and resources we did not modify the definition of image features and the neural network training, although these aspects may bear considerable optimization potential.

Unfortunately, the evaluation of our re-trained NFIQ+ algorithm only shows a slight improvement over the original NFIQ algorithm of NIST. As potential reasons, we suspect the internal thresholding of one SDK and a non-optimal neural network training. Nevertheless, a lot of experience was gained and lessons were learned regarding the statistical measurement of fingerprint image quality and neural network training. Moreover, several potentials for optimization were detected, which can be structured according to the 5 main steps in the general re-training process:

- *Fingerprint data basis:* The selection of the fingerprint data basis used for training is essential for the results of the re-training. Special data sets may lead to specialized behavior of the re-trained algorithm, which may or may not be desired (see end of this section for a discussion). Key factors of this area are (beside the number of fingerprints) the type of fingerprints (plain or rolled, inked or live-captured), the sensors used for capturing, and the number of imprints per finger.

- *Calculation of similarity score statistics:* The selection of the fingerprint SDKs (comprising feature extraction and comparison algorithms) used for the re-training will impact the applicability of the resulting NFIQ algorithm. As with the fingerprint data basis (see above) specialization for specific SDKs may or may not be desired. Furthermore, the definition of the normalized match scores determines the relevance of the NFIQ class definition to the actual application scenario.

- *Definition of NFIQ classes:* The number and (according to our evaluation, to a lower extent) the distribution of the NFIQ classes can have a significant influence on the results of the re-training. Furthermore, it is important to define how to merge similarity scores obtained from different SDKs.

- *Image feature selection:* We believe that the current definition of the image features used to predict the NFIQ leaves great optimization potential. However, the features must be chosen with utmost care to ensure that they are highly significant for the NFIQ. The number of features is also crucial as too few features may not reflect sufficient information about the image quality, and too many features may render training of a neural network difficult.

- *Neural network training:* Training the neural network may be strongly influenced by adapting and changing various parameters of the neural network. This is particularly true for a larger number of features. Key factors for successful neural network training are the size of the network, the error function but also other optimization methods and parameters described in Section 3.6.

Figure 3 illustrates the areas and key factors for optimizations.

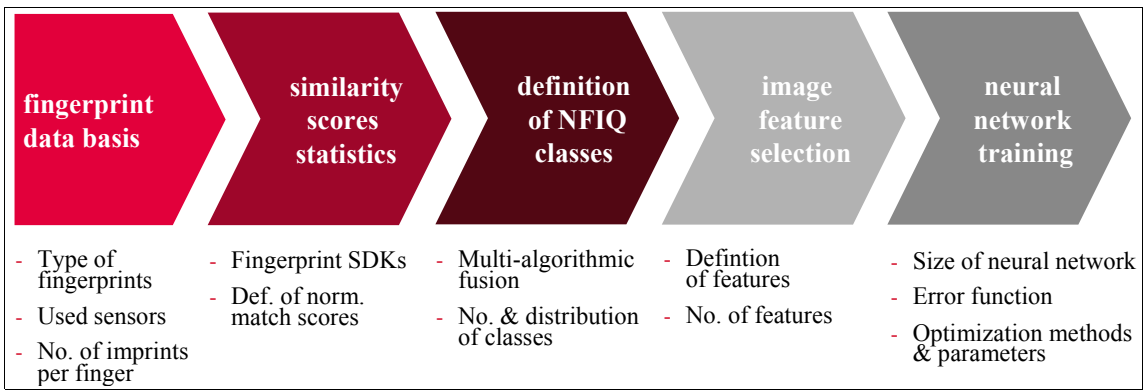

*Figure 3: Areas and key factors for optimization of the NFIQ.*

Summarizing, we are confident that a re-training of the NFIQ bears considerable potential for improvement.

For future attempts to optimize the NFIQ algorithm, we propose the following general two-step approach:

1. *Optimization of classification:* In a first step, the statistical classification, i.e. the NFIQ classes definition, should be optimized, as it constitutes the basis for the prediction by the neural network. This comprises the first three areas depicted in Figure 3, in particular, selection of the data basis and appropriate SDKs, calculation of the similarity score statistic, and its translation into NFIQ classes. The potential of a candidate classification can be evaluated by the DET curve of the perfect class predictor (see Section 4.3) as shown in Figure 2.

2. *Optimization of prediction:* In the second step, the most promising NFIQ classification identified in step 1 is used as a basis to develop a prediction algorithm. Within this step an eligible set of image features is selected, and a neural network is trained using various parameters and methods to achieve an optimal prediction accuracy.

Developing several specialized NFIQ algorithms is also an option. Specialized versions with respect to certain biometric sensors (or classes of them) or to certain fingerprint comparison algorithms may be useful in some application scenarios. Nevertheless, a generalized version of the NFIQ algorithm is still required. For instance, in case of electronic passports and identity cards, where fingerprint image data is stored for verification of the document holder at border control posts, such a generalized version of NFIQ is crucial.

We also would like to highlight that the plans proposed by NIST for the future of NFIQ are in line with our ideas. NIST suggests to use plug-and-play feature vectors in 3 different versions for a modular NFIQ 2.0 [T10]:

- *Black Box Feature Vector:* Proprietary quality feature vectors are completely deployed by third party vendors.
- *Clear Box Feature Vector:* A set of quality components is defined and maybe standardized. An open-source implementation will be developed by NIST.
- *Gray Box Feature Vector:* Combining standardized feature vectors (clear box) and vendor-dedicated feature vectors (black box) shall be another option.

In advance to further research in optimization of the NFIQ algorithm, we recommend to combine our ideas with the ideas proposed by NIST. In particular, a modular design of the new NFIQ 2.0 is highly appreciated. Despite having a dedicated version of NFIQ for certain scenarios there is still high demand for a generalized but improved version of the NFIQ algorithm. Mainly governmental authorities rely on an open-source implementation of NFIQ. Thus, for future enhancements of NFIQ, it is still essential to have a generalized and free version of the algorithm in order to be used within governmental fingerprint enrollment and authentication scenarios. We recommend to coordinate activities for the development of a future NFIQ 2.0 algorithm with NIST.